\documentclass{physeauth}
\usepackage{graphicx}
\usepackage{bm,amsmath}
\usepackage{amssymb}
\usepackage[numbers,comma,square,sort,sort&compress]{natbib}

\newcommand{\vecbm}[1]{\mbox{\boldmath#1}}

\newcommand{\nvec}[1]{\stackrel{\rightarrow}{#1}}
\newcommand{\stdef} {\stackrel{\mbox{\scriptsize def}}{=}}
\newcommand{\goo}{\,\raisebox{-.5ex}{$\stackrel{>}{\scriptstyle\sim}$}\,}
\begin{document}

%\catcode`\@=11
%\immediate\write\@auxout{\string\bibstyle{elsart-num}}  %  elsart-num
%\catcode`\@=12

\begin{frontmatter}

\title{Microcanonical Thermostatistics as Foundation of Thermodynamics.\\
 The microscopic origin of condensation and phase separations.}
\author{D.H.E. Gross}
\address{Hahn-Meitner Institut and Freie Universit{\"a}t Berlin,
Fachbereich Physik, Glienickerstr. 100, 14109 Berlin, Germany}
\thanks[thank1]{
Corresponding author. E-mail: gross@hmi.de}
\begin{abstract}
% Text of abstract
Conventional thermo-statistics address infinite homogeneous
systems within the canonical ensemble. However, some 150 years ago
the original motivation of thermodynamics was the description of
steam engines, i.e. boiling water. Its essential physics is the
separation of the gas phase from the liquid. Of course, boiling
water is inhomogeneous and as such cannot be treated by canonical
thermo-statistics. Then it is not astonishing, that a phase
transition of first order is signaled canonically by a Yang-Lee
singularity. Thus it is only treated correctly by microcanonical
Boltzmann-Planck statistics. This is elaborated in the present
article. It turns out that the Boltzmann-Planck statistics is much
richer and gives fundamental insight into statistical mechanics
and especially into entropy. This can even be done to some extend
rigorously and analytically. The microcanonical entropy has a very
simple physical meaning: It measures the microscopic uncertainty
that we have about the system, i.e. the number of points in
$6N$-dim phase, which are consistent with our information about
the system. It can rigorously be split into an ideal-gas part and
a configuration part which contains all the physics and especially
is responsible for all phase transitions. The deep and essential
difference between ``extensive'' and ``intensive'' control
parameters, i.e. microcanonical and canonical statistics, is
exemplified by rotating, self-gravitating systems.
\end{abstract}


\end{frontmatter}

% main text
\section{Introduction}
Since the beginning of Thermodynamics in the middle of the 19.century its
main motivation was the description of steam engines and the liquid to gas
transition of water. Here water prefers to become inhomogeneous and develop
a separation of the gas phase from the liquid, i.e. water boils. As
conventional canonical statistics works only for homogeneous, infinite
systems, phase separations remain outside of standard Boltzmann-Gibbs
thermo-statistics, which, consequently, signal phase-transitions of first
order by Yang-Lee singularities.

It is amusing that this fact that is essential for the original purpose of
Thermodynamics to describe steam engines was treated incompletely in the
past 150 years.  The system must be somewhat artificially split into (still
macroscopic) pieces for each individual phase \cite{guggenheim67}.

For this purpose, and also to describe small systems or non-extensive ones
like self-gravitating very large systems, we need a new and deeper
definition of statistics and as the heart of it: of entropy.

Also the second law can rigorously be formulated only microcanonically.
Already Clausius \cite{clausius1850,clausius1854,clausius1865,prigogine54}
distinguished between external and internal entropy generating mechanisms.
Canonical Boltzmann-Gibbs statistics is not sensitive to this important
difference.

\section{What is entropy?} Entropy, S, is the
fundamental entity of thermodynamics; therefore, its proper understanding
is essential. The understanding of entropy is sometimes obscured by
frequent use of the Boltzmann-Gibbs canonical ensemble, and the
thermodynamic limit. Also its relationship to the second law is often beset
with confusion between external transfers of entropy $dS_e$ and its
internal production $dS_i$.

The main source of the confusion is of course the lack of a clear {\em
microscopic and mechanical} understanding of the fundamental quantities of
thermodynamics like heat, external vs. internal work, temperature, and last
not least entropy, at the times of Clausius and possibly even today.

Clausius \cite{clausius1850,clausius1854} defined  a quantity which he
first called the ``{\em value of metamorphosis}'' in \cite{clausius1854}.
Eleven years later he \cite{clausius1865} gave it the name ``entropy'' $S$:
\begin{equation}
S_b-S_a=\int_a^b{\frac{dE}{T}},\label{entropy}
\end{equation} where $T$ is the absolute temperature of the body when the
momentary change is done, and $dE$ is the increment (positive
resp. negative) of all different forms of energy (heat and
potential) put into resp. taken out of the system. (Later,
however, we will learn that care must be taken of additional
constraints on other control parameters like e.g. the volume, see
below).

From the observation that heat does not flow from cold to hot (see section
\ref{zerolaw}, however section \ref{chsplit}) he went on to enunciate the
second law as:
\begin{equation}
\Delta S=\oint{\frac{dE}{T}}\ge 0,\label{secondlaw}
\end{equation}
which Clausius called the "{\em uncompensated metamorphosis}". As will be
worked out in section \ref{chsplit} the second law as presented by
eq.(\ref{secondlaw}) remains valid even in cases where heat flows from low
to higher temperatures.

Prigogine \cite{prigogine54}, c.f. \cite{guggenheim67}, quite clearly
stated that the variation of $S$ with time is determined by two distinct
mechanisms: the flow of entropy $dS_e$ to or from the system under
consideration; and its internal production $dS_i$. While the first type of
entropy change $dS_e$ (that effected by exchange of heat with its
surroundings) can be positive, negative or zero, the second type of entropy
change $dS_i$ (that caused by the internal creation of entropy) can be only
positive in any spontaneous transformation.

Clausius gives an illuminating example in \cite{clausius1854}: When an
ideal gas suddenly streams under insulating conditions from a small vessel
with volume $V_1$ into a larger one ($V_2>V_1$), neither its internal
energy $U$, nor its temperature changes, nor external work done, but its
internal (Boltzmann-)entropy $S_i$ rises, by $\Delta S=N\ln{(V_2/V_1)}$,
c.f. eq.(\ref{idgas}). Only by compressing the gas (e.g. isentropically)
and creating heat $\Delta E=E_1[(V_2/V_1)^{2/3}-1]$ (which must be finally
drained) it can be brought back into its initial state. Then, however, the
entropy change in the cycle, as expressed by integral (\ref{secondlaw}), is
positive ($=N\ln{(V_2/V_1)}$). This is also a clear example for a
microcanonical situation where the entropy change by an irreversible
metamorphosis of the system is absolutely internal. It occurs  during the
first part of the cycle, the expansion, where there is no heat exchange
with the environment, and consequently no contribution to the
integral(\ref{secondlaw}). The construction by eq.(\ref{secondlaw}) is
correct though artificial. After completing the cycle the Boltzmann-entropy
of the gas is of course the same as initially. All this will become much
more clear by Boltzmann's microscopic definition of entropy, which will
moreover clarify its real {\em statistical} nature:

Boltzmann\cite{boltzmann1877} later defined the entropy of an isolated
system (for which the energy exchange with the environment $dQ_e = 0$) in
terms of the sum of possible configurations, $W$, which the system can
assume consistent with its constraints of given energy  and
volume:\begin{equation} \fbox{\fbox{\vecbm{S=k*lnW}}}\label{boltzmann0}
\end{equation}as written on Boltzmann's tomb-stone, with
\begin{equation}
W(E,N,V)= \int{\frac{d^{3N}\nvec{p}\;d^{3N}\nvec{q}}{N!(2\pi\hbar)^{3N}}
\epsilon_0\;\delta(E-H\{\nvec{q},\nvec{p}\})}\label{boltzmann}
\end{equation} in semi-classical approximation. $E$ is the total energy, $N$ is
the number of particles and $V$ the volume. Or, more appropriate for a
finite quantum-mechanical system:
\begin{eqnarray} W(E,N,V)&=& Tr[\mathcal{P}_E]\label{quantumS}\\
&=&\sum{\scriptsize\begin{array}{ll}\mbox{all eigenstates n of H with given
N,$V$,}\\\mbox{and } E<E_n\le E+\epsilon_0\nonumber
\end{array}}
\end{eqnarray}
and $\epsilon_0\approx$ the macroscopic energy resolution.  This is still
up to day the deepest, most fundamental, and most simple definition of
entropy. There is no need of the thermodynamic limit, no need of concavity,
extensivity and homogeneity.  In its semi-classical approximation,
eq.(\ref{boltzmann}), $W(E,N,V,\cdots)$ simply measures the area of the
sub-manifold of points in the $6N$-dimensional phase-space ($\Gamma$-space)
with prescribed energy $E$, particle number $N$, volume $V$, and some other
time invariant constraints which are here suppressed for simplicity.
Because it was Planck who coined it in this mathematical form, I will call
it the Boltzmann-Planck principle.

There are various reviews on the mathematical foundations of statistical
mechanics, e.g., the detailed and instructive article by Alfred
Wehrl\cite{wehrl78}. Wehrl shows how the Boltzmann-Planck formulae,
equations (\ref{boltzmann0}) and (\ref{quantumS}), may be generalized to
the famous definition of entropy in quantum mechanics by von Neumann
\cite{neumann27}:
\begin{equation}S=-Tr[\rho\ln(\rho)],\label{neumann}
\end{equation}addressing general (also non projector like) densities $\rho$.
Wehrl discusses the conventional, canonical, Boltzmann-Gibbs statistics
where all constraints on $\rho$ are fixed only to their mean, allowing for
free fluctuations. These free, unrestricted fluctuations of the energy
$\Delta E=\sqrt{<(E-<E>)^2>}$ imply an uncontrolled energy exchange with
the universe, $dQ_e$ in Prigogine's definition. For the homogeneous phase
of a system with short-ranged interactions $\frac{\Delta E}{<E>}$ vanishes
in the thermodynamic limit. In general this, however, is dangerous; for
there are situations where the fluctuations are macroscopic and
$\frac{\Delta E}{<E>}$ does not vanish in the thermodynamic limit. An
example are phase transitions of first oder where $\Delta E = E_{latent}$,
the latent heat of transformation. Another well known is the case of
long-ranged interactions like in gravitating systems. Wehrl points to many
serious complications with this definition.

However, in the case of conserved variables, we know more than their mean;
we know these quantities sharply. In microcanonical thermodynamics, we do
not need von Neumann's definition (\ref{neumann}), and can work on the
level of the original, Boltzmann-Planck definition of entropy, equations
(\ref{boltzmann0}) and (\ref{quantumS}). We thus explore statistical
mechanics and entropy at their most fundamental level. This has the great
advantage that the axiomatic level is extremely simple. Because such
analysis does not demand scaling or extensivity, it can further be applied
to the much wider group of non-extensive systems from nuclei to galaxies
and address the original object for which thermodynamics was enunciated
some 150 years ago: phase separations.

The Boltzmann-Planck formula has a simple but deep physical interpretation:
$W$ or $S$ are the measure of our ignorance about the complete set of
initial values for all $6N$ microscopic degrees of freedom which are needed
to specify the $N$-body system unambiguously\cite{kilpatrick67}. To have
complete knowledge of the system we would need to know (within its
semiclassical approximation (\ref{boltzmann})) the initial position and
velocity of all $N$ particles in the system, which means we would need to
know a total of $6N$ values. Then $W$ would be equal to one and the
entropy, $S$, would be zero. However, we usually only know the value of a
few parameters that change slowly with time, such as the energy, number of
particles, volume and so on. We generally know very little about the
positions and velocities of the particles. The manifold of all these points
in the $6N$-dim. phase space is the microcanonical ensemble, which has a
well-defined geometrical size $W$ and, by equation (\ref{boltzmann0}), a
non-vanishing entropy, $S(E,N,V,\cdots)$. The dependence of
$S(E,N,V,\cdots)$ on its arguments determines completely thermostatics and
equilibrium thermodynamics.

Clearly, Hamiltonian (Liouvillean) dynamics of the system cannot create the
missing information about the initial values, - i.e., the entropy,
$S(E,N,V,\cdots)$ cannot decrease. As has been further worked out
\cite{gross183} and more recently in \cite{gross207} the inherent finite
resolution of the macroscopic description implies an increase of $W$ or $S$
with time when an external constraint is relaxed. Such is a statement of
the second law of thermodynamics, which requires that the internal
production of entropy be positive for every spontaneous process. Analysis
of the consequences of the second law by the microcanonical ensemble is
appropriate because, in an isolated system (which is the one relevant for
the microcanonical ensemble), the changes in total entropy must represent
the internal production of entropy, and there are no additional
uncontrolled fluctuating energy exchanges with the environment.

\section{The Zero'th Law in conventional extensive Thermodynamics\label{zerolaw}}
This section and the following discuss mainly systems that have no other
macroscopic (extensive) control parameter besides energy; the particle
density is not changed, and there are no chemical reactions.

In conventional (extensive) thermodynamics thermal equilibrium of two
systems (1 \& 2) is established by bringing them into thermal contact which
allows free energy exchange. Equilibrium is established when the total
entropy
\begin{equation}
S_{total}(E,E_1)=S_1(E_1)+S_2(E-E_1)\label{eq1}
\end{equation}
is maximal:
\begin{equation}
dS_{total}(E,E_1)|_E=dS_1(E_1)+dS_2(E-E_1)=0\label{eq2}.
\end{equation}
Under an energy flux $\Delta E_{2\to 1}$ from $2\to 1$ the total entropy
changes to lowest order in $\Delta E$ by
\begin{eqnarray}
\Delta S_{total}|_E&=&(\beta_1-\beta_2)\Delta E_{2\to 1}\\
\beta&=&dS/dE=\frac{1}{T}
\end{eqnarray}
Consequently, a maximum of $S_{total}(E=E_1+E_2,E_1)|_E$ will be approached
when
\begin{equation}
 \mbox{sign}(\Delta S_{total})=\mbox{sign}(T_2-T_1)\mbox{sign}(\Delta E_{2\to 1})>0
\end{equation}
From here Clausius' first formulation of the second law follows:
"Heat always flows from hot to cold". Essential for this
conclusion is the {\em additivity} of $S$ under the split
(eq.\ref{eq1}). There are no correlations, which are destroyed
when an extensive system is split. Temperature is an appropriate
control parameter for extensive systems.

\section{No phase separation without a convex, non-extensive $S(E)$\label{chsplit}}

The weight $e^{S(E)-E/T}$ of the configurations with energy E in the
definition of the canonical partition sum
\begin{equation}
Z(T)=\int_0^\infty{e^{S(E)-E/T}dE}\label{canonicweight}
\end{equation} becomes here {\em bimodal}, at the transition temperature it has
two peaks, the liquid and the gas configurations which are separated in
energy by the latent heat. Consequently $S(E)$ must be convex and the
weight in (\ref{canonicweight}) has a minimum between the two pure phases.
Of course, the minimum can only be seen in the microcanonical ensemble
where the energy is controlled and its fluctuations forbidden. Otherwise,
the system would fluctuate between the two pure phases by an, for
macroscopic systems even macroscopic, energy $\Delta E\sim E_{lat}$ of the
order of the latent heat. I.e. {\em the convexity of $S(E)$ is the generic
signal of a phase transition of first order} and of
phase-separation\cite{gross174}.  Such macroscopic energy fluctuations and
the resulting negative specific heat are already early discussed in
high-energy physics by Carlitz \cite{carlitz72}.

\begin{figure}[h]
\includegraphics*[bb =37 1 453 357, angle=-90, width=9 cm,
clip=true]{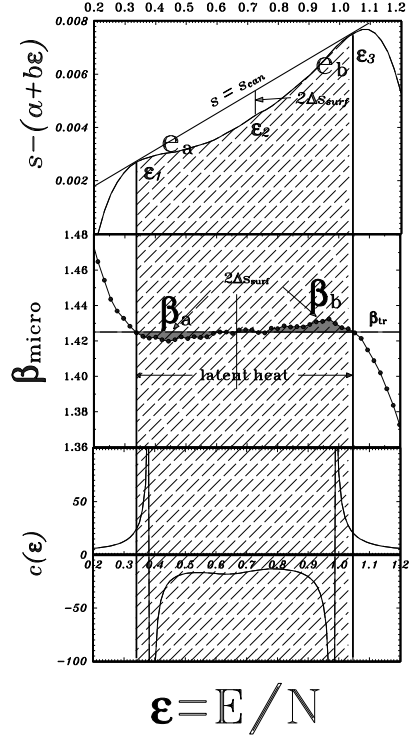}%{pr31AB.eps}
\caption{Ferromagnetic Potts model ($q=10$) on a $50*50$-lattice
with periodic boundary conditions in the region of phase
separation. At the energy $e_1$ per lattice point the system is in
the pure ordered phase, at $e_3$ in the pure disordered phase. At
$e_a$ little above $e_1$ the temperature $T_a=1/\beta$ is higher
than $T_2$ and even more than $T_b$ at $e_b$ a little below $e_3$.
At $e_a$ the system separates into a few bubbles of disordered
phase embedded in the ordered phase or at $e_b$ into a few
droplets of ordered phase within the disordered one. If we combine
two equal systems: one with the energy per lattice site
$e_a=e_1+\Delta e$ and at the temperature $T_a$, the other with
the energy $e_b=e_3-\Delta e$ and at the temperature $T_b<T_a$,
and allowing for free energy exchange, then the systems will
equilibrize at energy $e_2$ with a {\em rise} of its entropy. The
temperature $T_a$ drops (cooling) and energy (heat) flows (on
average) from $b\to a$. I.e.: {\em Heat flows from cold to hot!
Thus, the Clausius formulation of the second law is violated}.
This is well known for self-gravitating systems. However, this is
not a peculiarity of only gravitating systems! It is the generic
situation at phase separations within classical thermodynamics
even for systems with short-range coupling and has nothing to do
with long-range interactions.\label{figure 1}}
%%%\end{center}
\end{figure}

The ferromagnetic Potts-model illuminates in a most simple example the
occurrence of a convex intruder in $S(E)$ which induces a backbending
caloric curve $T(E)=(\partial S/\partial E)^{-1}$ with a decrease of the
temperature $T(E)$ with rising energy \cite{gross150}. A typical plot of
$s(e,N)=S(E=Ne)/N$ in the region of phase separation is shown in
fig(\ref{figure 1}). Section \ref{convex} discusses the general microscopic
reasons for the convexity.(Moretto et al\cite{moretto02} have
previously put forward errors connected with the use of periodic boundary
conditions; these assertions been rebutted.\cite{gross196,gulminelli03})

This has far reaching consequences which are crucial for the fundamental
understanding of thermo-statistics and Thermodynamics: Let us split the
system of figure (\ref{figure 1}) into two pieces $a$ \& $b$ by a dividing
surface, with half the number of particles each. The dividing surface is
purely geometrical. It exists only as long as the two pieces can be
distinguished by their different energy/particle $e_a$ and $e_b$.
Constraining the energy-difference $e_b-e_a=\Delta e$ between the two,
reduces the number of free, unconstrained degrees of freedom and {\em
reduces} the entropy by $-2\Delta S_{surf-corr.}$. (Moreover, if the effect
of the new surface would also be to cut some bonds: before the split there
were configurations with attractive interactions across the surface which
are interrupted by the division, their energy shifts upwards outside the
permitted band-width $\epsilon_0$, and thrown out of the partition sum
(\ref{quantumS}). I.e. the entropy will be further reduced by the split.)

If the constraint on the difference $e_b-e_a$ is fully relaxed and
$e_b-e_a$ can fluctuate freely at fixed $e_2=(e_a+e_b)/2$, the dividing
surface is assumed to have no further physical effect on the system.

\begin{figure}[h]
\includegraphics*[bb =0 40 467 323,width=6cm,angle=-0,clip=true]{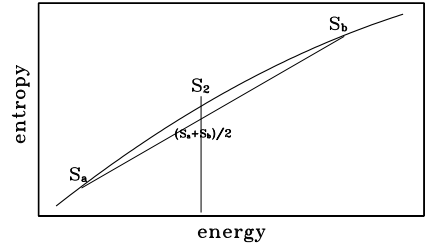}
\caption {Extensive (concave) $S(E)$\label{figure 2}}
\end{figure}

For an {\em extensive} system [$S(E,N)=N  s(e=E/N)=2S(E/2,N/2)$]. One would
argue as follows: The combination of two pieces of $N/2$ particles each,
one at $e_a=e_2-\Delta e/2$ and a second at $e_b=e_2+\Delta e/2$, must lead
to $S(E_2,N)\ge S(E_a/2,N/2)+S(E_b/2,N/2)$, the simple algebraic sum of the
individual entropies because by combining the two pieces one normally
looses information. This, however, is equal to $[S(E_a,N)+S(E_b,N)]/2$,
thus $S(E_2,N)\ge[S(E_a,N)+S(E_b,N)]/2$. I.e. {\em the entropy $S(E,N)$ of
an extensive system is necessarily concave}, c.f. figure(\ref{figure 2}).

For a {\em non-extensive} system we have in general $S(E,N)\ge 2S(E/2,N/2)$
because again two separated, closed pieces have more information than their
unification. Now, if $E_2$ is the point of maximum positive curvature of
$S(E,N)$ (convexity = upwards concave like $y=x^2$) we have $S(E_2,N) \le
[S(E_a,N)+S(E_b,N)]/2$ like in fig.(\ref{figure 1}). However, the r.h.s. is
larger than $S(E_a/2,N/2)+S(E_b/2,N/2)$. I.e. even though $S(E,N)$ is
convex at const. $N$, the unification of the pieces with $E_a/2,N/2$ and
$E_b/2,N/2$ can still lead to a {\em larger} entropy $S(E_2,N)$.

The difference between $[S(E_a,N)+S(E_b,N)]/2$ and
$S(E_a/2,N/2)+S(E_b/2,N/2)$ we call henceforth $\Delta S_{surf-corr}$. The
correct entropy balance, before and after establishing the energetic split
$e_b>e_a$ of the system, is
\begin{equation}
 S_{after}-S_{before}=\frac{S_a+S_b}{2}-\Delta S_{surf-corr.}-S_2 \le
 0 \label{balance}
\end{equation} even though the difference of the first and the last term is positive.

In the inverse direction: By {\em relaxing} the constraint and allowing, on
average, for an energy-flux ($\Delta E_{b\to a}>0$) {\em opposite to
$T_a-T_b>0$, against the temperature-gradient (slope)}, but in the
direction of the energy-slope, the entropy $S_{total}\to S_2$ increases.
This is consistent with the naive picture of an {\em energy equilibration}.
Thus {\em Clausius' "energy flows always from hot to cold", i.e. the
dominant control-role of the temperature in thermo-statistics
\cite{hertz10a} is violated}. Of course this shows again that {\em unlike
to extensive thermodynamics the temperature is not the appropriate control
parameter in non-extensive systems.}

In the thermodynamic limit $N\to\infty$ of a system with short-range
coupling $\Delta S_{surf-corr.}\sim N^{2/3}$, $\Delta
S_{surf-corr.}/N=\Delta s_{surf-corr.}\propto N^{-1/3}$ must go to $0$ due
to van Hove's theorem.

\section{The origin of the convexities of $S(E)$ and of phase-separation\label{convex}} Many
applications of microcanonical thermodynamics to realistic examples of hot
nuclei, atomic clusters, and rotating astrophysical systems have been
presented during the past twenty years which demonstrate convex intruders
in the microcanonical entropy and, consequently, negative heat capacities.
Such are reviewed in the publication list on the web site
http://www.hmi.de/people/gross/ and
elsewhere\cite{chomaz99,chomaz00,chomaz00a}. Here we shall illuminate the
general microscopic mechanism leading to the appearance of a convex
intruder in $S(E,V,N,,\cdots)$ as far as possible by rigorous and
analytical methods. This is the generic signal of phase transitions of
first order and of phase-separation within the microcanonical ensemble.
Assume the system is classical and obeys the Hamiltonian:
\begin{eqnarray}H &= &\sum_i^N{\frac{p_i^2}{2m}}+\Phi^{int}[\{\nvec{r}\}]
\label{hamiltonian}\\\nonumber\\\nonumber\\
\Phi^{int}[\{\nvec{r}\}]&:=&\sum_{i<j}{\phi(\nvec{r}_i-\nvec{r}_j)}\nonumber
\end{eqnarray}
 In this case the system is
controlled by energy and volume.

\subsection{Liquid-gas transition} The microcanonical sum of states or partition sum is:

\begin{eqnarray}
W(E,N,V)&=&\frac{1}{N!(2\pi\hbar)^{3N}}\times\label{W}\\
\lefteqn{\hspace{-3cm}\int_{V^N}{d^{3N}\nvec{r} \int{d^{3N}\nvec{p}_i{
\epsilon_0\;\delta(E-\sum_i^N{\frac{\nvec{p}_i^2}{2m_i}-\Phi^{int}[\{\nvec{r}\}]})}}}}
\nonumber\\\nonumber\\&=&\nonumber\\\nonumber\\\lefteqn{\hspace{-2cm}\frac{V^N
\epsilon_0(E-E_0)^{(3N-2)/2} \prod_1^N{m_i^{3/2}}}{N!\Gamma(3N/2)
(2\pi\hbar^2)^{3N/2}}\;\;\times\nonumber}\\
\lefteqn{\hspace{-2cm}\int_{V^N}{\frac{d^{3N}r}
{V^N}}\left(\frac{E-\Phi^{int}[\{\nvec{r}\}]}{E-E_0}\right)^{(3N-2)/2}
\label{split}}\\\nonumber\\
&=&\nonumber\\\lefteqn{\hspace{-3cm}W_{id-gas}(E-E_0,N,V)\times
W_{conf}(E-E_0,N,V)\nonumber}\\\nonumber\\&=&
e^{[S_{id-gas}+S_{conf}]}\label{micromeg}\\\nonumber\\
W_{id-gas}(E,N,V)&=&\frac{V^N\epsilon_0
E^{(3N-2)/2}\prod_1^N{m_i^{3/2}}}{N!\Gamma(3N/2)
(2\pi\hbar^2)^{3N/2}}\label{idgas}\\\nonumber\\\nonumber\\
W_{conf}(E-E_0,N,V)&=&\int_{V^N}{\frac{d^{3N}r}
{V^N}}\Theta(E-\Phi^{int}[\{\nvec{r}\}])\nonumber\\ \lefteqn{\hspace{-1cm}
\times\left(1-\frac{\Phi^{int}[\{\nvec{r}\}]-E_0}{E-E_0}\right)^{(3N-2)/2}
\label{Win1}}
\end{eqnarray}
V is the spatial volume; $E_0=\min \Phi^{int}[\{\nvec{r}\}]$ is the energy
of the ground-state of the system. The separation of $W(E,N,V)$ into
$W_{id-gas}$ and $W_{conf}$ is the microcanonical analogue of the split of
the canonical partition sum into a kinetic part and a configuration part:
\begin{equation}
Z(T)=\frac{V^N}{N!}\left(\frac{m
T}{2\pi\hbar^2}\right)^{3N/2}\int{\frac{d^{3N}r}{V^N}e^{-\frac{\Phi^{int}[\{\nvec{r}\}]}{T}}}\label{canonical}
\end{equation}

In the thermodynamic limit, the order parameter of the
(homogeneous) liquid-gas transition is the density. The transition
is linked to a condensation of the system towards a larger density
controlled by pressure. For a finite system, we expect analogous
behavior. However, for a closed finite system, which is allowed to
become inhomogeneous at phase separation, this is controlled by
the available system volume $V$ and not by intensive density or
pressure. At low energies, the $N$ particles condensate into a
droplet with much smaller volume $V_{0,N}\ll V$. $3(N-1)$ internal
coordinates are limited to $V_{0,N}$. Only the center of mass of
the droplet can move freely in $V$ (remember we did not fix the
center-of-mass in equation eq.(\ref{W})). The system does not fill
the $3N$-configuration space $V_N$. Only a stripe with width $
V_{0N}^{1/3}$ in $3(N-1)$ dimensions of the total $3N$-dim space
is populated. The system is non-homogeneous even though it is
equilibrized and, at low energies, internally in the single liquid
phase; and it is not characterized by an intensive homogeneous
density. In fact, $W_{conf}(E-E_0,N,V)$ can be written as:
\begin{eqnarray}
W_{conf}(E-E_0,N,V)&=& \left[\frac{V(E,N)}V\right]^N\le 1 \label{Win1b}\\
\left[V(E,N)\right]^N&\stdef&\nonumber\\
 \lefteqn{\hspace{-3cm}\int_{V^N}d^{3N}r\;\Theta(E-\Phi^{int}[\{\nvec{r}\}])}\nonumber\\
\lefteqn{\hspace{-3cm}\times\left(1-\frac{\Phi^{int}[\{\nvec{r}\}]-E_0}{E-E_0}\right)^{(3N-2)/2}}
\label{Sin2}\\\nonumber\\
S_{conf}(E-E_0,N,V)&=&N\ln\left[\frac{V(E,N)}{V}\right]\le0\label{Sin1}
\end{eqnarray}
The first factor $\Theta(E-\Phi^{int}[\{\nvec{r}\}])$ in eq(\ref{Sin2})
eliminates the energetically forbidden regions. Only the potential holes
(clusters) in the $3N$-dim potential surface $\Phi^{int}[\{r\}]\le E$
remain. Their volume $V^N(E,N)\le V^N$ is the accessible part of the
$3N$-dim-spatial volume where $\Phi^{int}[\{r\}]\le E$. I.e. $ V^N(E,N)$ is
the total $3N$-dim. eigen-volume of the condensate (droplets), with $N$
particles at the given energy, summed over all possible partitions,
clusterings, in $3N$-configuration space. The relative volume fraction of
each partition compared with $V^N(E,N)$ gives its relative probability. $
V^N(E,N)$ has the limiting values:
\begin{equation}
{[V(E,N)]^N}=\left\{\begin{array}{ll}V^N&\mbox{for $E$ in the gas
phase}\nonumber\\{V_{0N}}^{N-1}V&\mbox{for~~}E=E_0
\end{array}\right.\label{volume}\end{equation}
$W_{conf}(E-E_0,N,V)$ and $S_{conf}(E-E_0,N,V)$ have the limiting values:
\begin{eqnarray} W_{conf}(E-E_0)&\le& 1,\;\Rightarrow S_{conf}(E-E_0,N)
 \le0\nonumber\\
&\rightarrow& \left\{\begin{array}{ll}1 &\;\;\; \;\;\;\;\;\; \;\;\;E\gg
\Phi^{int}\nonumber\\\left[\frac{V_{0N}}V\right]^{(N-1)}&\;\;\;
\;\;\;\;\;\; \;\;\;E\to E_0\end{array}\right.
\\\\\nonumber\\ S_{conf}(E-E_0)&\to
&\left\{\begin{array}{ll}0&E\gg
\Phi^{int}\nonumber\\ln\left\{[\frac{V_{0N}}V]^{N-1}\right\}< 0&E\to
E_0\end{array}\right. \label{Sin2b}\\\end{eqnarray}

All physical details are contained in $W_{conf}(E-E_0,N,V)$ alias
$N\ln[V(E,N)]$, c.f. eqs.(\ref{Win1b}--\ref{Sin2b}): If the energy is high
the detailed structure of $\Phi^{int}[\{\nvec{r}\}]$ is unimportant
$W_{conf}\approx 1$, $S_{conf}\approx 0$. The system behaves like an ideal
gas and fills the volume $V$. At sufficiently low energies only the minimum
of $\Phi^{int}[\{\nvec{r}\}]$ is explored by $W_{conf}(E-E_0,N, V)$. The
system is in a condensed phase, a single liquid drop, which moves freely
inside the empty larger volume $V$, the $3(N-1)$ internal degrees of
freedom are trapped inside the {\em reduced} volume $V_{0N} \ll V$.

One can guess the general form of $N\ln[V(E,N)]$: Near the groundstate
$E\goo E_0$ it must be flat $\approx(N-1)\ln[V_{0N}]+\ln[V-V_{0N}]$ because
the liquid drop has some eigen-volume $V_{0N}$  in which each particle can
move (liquid). With rising energy $\ln[ V(E,N)]$ rises up to the point
($E_{trans}$) where it is possible that the drop fissions into two. Here an
additional new configuration opens in $3N$-dim configuration space: Either
one particle evaporates from the cluster and explores the external volume $
V$, or the droplet fissions into two droplets and the two CM coordinates
explore the larger $V$. This gives a sudden jump in $S_{conf}(E)$ by
something like $\sim \ln\{\frac{V-V_{0(N-1)}}{V_{0(N-1)}}\}$ and similar
jump upwards in the second case.

Later further such "jumps" may follow. Each of these "jumps" induce a
convex upwards bending of the total entropy $S(E)$ (eq.\ref{micromeg}).
Each is connected to a bifurcation and bimodality of $e^{S(E)-E/T}$ and the
phenomenon of {\em phase-separation}.

In the conventional canonical picture for a large number of particles this
is forbidden and hidden behind the familiar Yang-Lee singularity of the
liquid to gas phase transition. In the microcanonical ensemble this is
analogue to the phenomenon of multi-fragmentation in nuclear systems
\cite{gross174,gross153}. This, in contrast to the mathematical Yang-Lee
theorem, physical microscopic explanation of the liquid to gas phase
transition sheds sharp light on the physical origin of the transition, the
sudden change in the inhomogeneous population of the $3N$-dim.
configuration space.

\subsection{Solid-liquid transition} In contrast to the liquid
phase, in the crystal phase a molecule can only move locally within its
lattice cage of the size $d^3$ instead of the whole volume $V_{0N}$ of the
condensate. I.e. in equation (\ref{Sin2b}) instead we have
$S_{conf}\to\ln\{[\frac{d^3}{V_{0N}}]^{N-1}\}$.

\subsection{Summary of section V}The essential differences between
the gas, the liquid, and solid phase are the following: Whereas the gas
occupies the whole container, the liquid is confined to a definite
condensate volume, however this may have any shape. It is separated from
the gas by a surface. The solid is also confined to definite volume but in
contrast to the liquid its surface has also a definite shape. These
differences cannot be seen in the canonical ensemble.

The gas- liquid transition is linked to the transition from uniform filling
of the container volume $V$ by the gas to the smaller eigen-volume of the
system $V_0$ in its condensed phase where the system is inhomogeneous (some
liquid drops inside the larger empty volume $V$). First $3(N-1)$, later at
higher energies less and less degrees of freedom condensate into the drop.
First three, then more and more degrees of freedom
(center-of-mass-coordinates of the drops) explore the larger container
volume $V$ leading to upwards jumps (convexities) of $S_{conf}(E$). The
volume of the container controls how close one is to the critical end-point
of the transition, where phase-separation disappears. Towards the critical
end-point, i.e. with smaller V, the jumps $\ln[V-V_0]-\ln[V_0]$ become
smaller and smaller. In the case of the solid-liquid transition, however,
the external volume, $V$, of the container confines only the center-of-mass
position of the crystal, resp., the droplet. The entropy jumps during
melting by $\Delta S_{conf}\propto\ln[V_{0N}]-\ln{d^3}$. At the surface of
a drop $ \Phi^{int}> E_0=\min \Phi^{int}$, i.e. the surface gives a
negative contribution to $S_{conf}$ in equation (\ref{Sin2}) and to $S$ at
energies $E\goo E_0$, as was similarly assumed in section (\ref{chsplit})
and explicitly in equation (\ref{balance}).
\section{Application in astrophysics}
The necessity of using ``extensive'' instead of ``intensive'' control
parameter is explicit in astrophysical problems. E.g.: for the description
of rotating stars one conventionally works at a given temperature  and
fixed angular velocity $\Omega$ c.f. \cite{chavanis03}. Of course in
reality there is neither a heat bath nor a rotating disk. Moreover, the
latter scenario is fundamentally wrong as at the periphery of the disk the
rotational velocity may even become larger than velocity of light.
Non-extensive systems like astro-physical ones do not allow a
``field-theoretical'' description controlled by intensive fields !

E.g. configurations with a maximum of random energy
\begin{equation}
E_{random}=E-\frac{\Theta\Omega^2}{2} -E_{pot}
\end{equation} and consequently with the largest entropy are the ones
with smallest moment of inertia $\Theta$, compact single stars. Just the
opposite happens when the angular-momentum $L$ and not the angular velocity
$\Omega$ are fixed:\begin{equation} E_{random}=E-\frac{L^2}{2 \Theta}
-E_{pot}.
\end{equation}Then configurations with large moment of inertia are
maximizing the phase space and the entropy. I.e. eventually double or multi
stars are produced, as observed in reality.

\begin{figure}[h]
\hspace*{-0.5cm}
\includegraphics[bb =0 0 511 353,width=8cm,angle=0,clip=true]{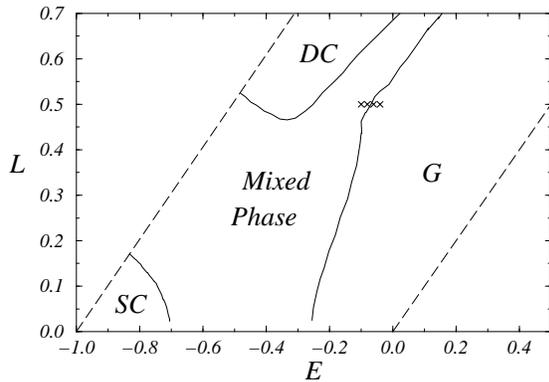}
\caption{Phase diagram of rotating self-gravitating systems in the
energy-angular-momentum $(E,L)$-plane. DC: region of double-stars, G: gas
phase, SC: single stars. In the mixed region one finds various exotic
configurations like ring-systems in coexistence with gas, double stars or
single stars. In this region of phase-separation the heat capacity is
negative and the entropy is convex. The dashed lines $E-L=1$ (left) and
$E=L$ (right) delimit the region where calculations were carried
out.\label{phased}}
\end{figure}

In figure \ref{phased} one clearly sees the rich and realistic
microcanonical phase-diagram of a rotating gravitating system controlled by
the ``extensive'' parameters energy and angular-momentum. \cite{gross187}

\section{Acknowledgement} I am grateful to J.F. Kenney for numerous
suggestions and to J. M\"oller for insistent, therefore helpful,
discussions.
%%%\bibliographystyle{unsrt}%{alpha}%{plain} %{unsrt}
%%%\bibliography{gross,othbiba,othbibb,othbibcd,othbibe,othbibf,othbibg,othbibh,othbibij,othbibk,othbibl,othbibm,othbibn,othbibo,othbibp,othbibr,othbibs,othbibt,othbibuw,othbibxz}
%%%\end{document}

\end{document}